\documentclass[12pt]{article}
\usepackage[english,german,french,polish]{babel}
\usepackage[T1]{fontenc}
\usepackage{amsfonts}

\begin{document}

\selectlanguage{english}

\baselineskip 0.73cm
\topmargin -0.4in
\oddsidemargin -0.1in

\let\ni=\noindent

\renewcommand{\thefootnote}{\fnsymbol{footnote}}

\newcommand{\SM}{Standard Model }

\pagestyle {plain}

\setcounter{page}{1}



~~~~~~
\pagestyle{empty}

\begin{flushright}
IFT-- 08/7
\end{flushright}

\vspace{0.4cm}

{\large\centerline{\bf Quasi-magnetism of sterile dark matter}}

{\large\centerline{\bf with photonic portal}}

\vspace{0.5cm} 

{\centerline {\sc Wojciech Kr\'{o}likowski}}

\vspace{0.3cm}

{\centerline {\it Institute of Theoretical Physics, University of Warsaw}}

{\centerline {\it Ho\.{z}a 69, 00--681 Warszawa, ~Poland}}

\vspace{0.6cm}

{\centerline{\bf Abstract}}

\vspace{0.2cm}

\begin{small}

\begin{quotation}

We develop farther a simple model of cold dark matter consisting of sterile {spin-1/2} fermions (sterinos), whose mass is generated by a nonzero vacuum expectation value of a field of sterile scalars (sterons). In this model,
the sterile world of sterinos and sterons  is conjectured to communicate with the familiar Standard Model world 
not only through gravity, but also through a photonic portal provided by a very weak effective interaction of 
the electromagnetic field with sterino and steron fields. Then, an extra, finite renormalization leading from the
primary bare electric charge to a new bare electric charge appears spontaneously. Due to the photonic portal, the cold dark matter  consisting of sterinos displays a phenomenon of quasi-magnetism, what means that it can interact effectively with cosmic (and laboratory) magnetic fields.
 
\vspace{0.6cm}

\ni PACS numbers: 14.80.-j , 04.50.+h , 95.35.+d 

\vspace{0.3cm}

\ni May 2008

\end{quotation}
 
\end{small}

\vfill\eject

\pagestyle {plain}

\setcounter{page}{1}

\vspace{0.3cm}

\ni {\bf 1. Introduction}

\vspace{0.3cm} 

In a recent work [1], a simple model for cold dark matter consisting of spin-1/2 fermions, sterile from \SM charges, has been proposed, where masses are generated by a nonzero vacuum expectation value of a field of scalar bosons, also assumed to be sterile. It has been conjectured that --- after the electroweak symmetry is broken and related photons emerge --- the fields $\psi$ and $\varphi$  of these sterile particles interact directly with the electromagnetic field $F_{\mu\,\nu} = \partial_\mu A_\nu - \partial_\nu A_\mu $ through the very weak effective coupling

\vspace{-0,1cm}

\begin{equation}
-\frac{f}{M^2}\,F_{\mu\,\nu}\,\varphi F^{\mu\,\nu}\,\varphi - \frac{f'}{M^2}\, (\bar\psi \sigma_{\mu\,\nu} \psi )\, F^{\mu\,\nu}\varphi \;.
\end{equation}

\ni Here, $M$ is a very large mass scale, while $f$ and $f'$ are unknown dimensionless coupling constants. This mechanism has been called the photonic portal (to the sterile world). It is an alternative for the popular Higgs portal (to the sterile world) provided by a direct interaction of some sterile scalars with the \SM Higgs bosons 
({\it cf.} Ref. [2] for a recent discussion including Sommerfeld corrections to this interaction).

Beside the effective coupling (1), our $\psi$ and $\varphi$ fields interact gravitationally as well as participate in the mass-generating Higgs-type coupling

\begin{equation}
-(\bar{\psi}\,Y\, \psi) \varphi + \frac{1}{2}\mu^2 \varphi^2 - \frac{1}{4}\lambda \varphi^4\;\;\;{\rm with}\;\;\; <\!\!\varphi\!\!>_{\rm vac}\, \neq 0 \,,
\end{equation}

\ni where $Y$ is a constant matrix if there are more generations of $\psi $ fermions. Here, the physical scalar field is $\varphi_{\rm ph} \equiv \varphi\, - <\!\!\varphi\!\!>_{\rm vac}$.  For convenience in their presentation, the $\psi$ and $\varphi$ sterile particles have been called {\it sterinos} and {\it sterons}, respectively

\vspace{0.3cm}

\ni {\bf 2. Spontaneous renormalization}

\vspace{0.3cm} 

When the effective interaction (1) is added to the electromagnetic Lagrangian $-\frac{1}{4} F_{\mu\,\nu} F^{\mu\,\nu}- j_\mu A^\mu $, where $F_{\mu\,\nu} = \partial_\mu A_\nu - \partial_\nu A_\mu $ and $j_\mu$ is the \SM
electromagnetic current, the resulting effective Lagrangian is

\begin{equation}
{\cal{ L }} = -\frac{1}{4} F_{\mu\,\nu} F^{\mu\,\nu} \left(1+ \frac{4f}{M^2} \varphi^2 \right) - \frac{f'}{M^2}(\bar\psi \sigma_{\mu\,\nu} \psi) F^{\mu\,\nu} \varphi - j_\mu A^\mu\;. 
\end{equation}

\ni With $\varphi \equiv <\!\!\varphi\!\!>_{\rm vac} + \varphi_{\rm ph}$, the first term in Eq. (3) can be rewritten as

\begin{equation}
-\frac{1}{4} F_{\mu\,\nu} F^{\mu\,\nu} Z^{-1}\left[ 1+ \frac{4f}{M^2}\left(2<\!\!\varphi\!\!>_{\rm vac}\varphi_{\rm ph} + \varphi^2_{\rm ph} \right)Z\right] \,,
\end{equation}

\ni where 

\begin{equation}
 Z^{-1} \equiv 1+ \frac{4f}{M^2}\!\!<\!\!\varphi\!\!>^2_{\rm vac}\;\; > 1 \;.
\end{equation}

Then, operating with the constant $Z^{-1}$, we can observe that the effective Lagrangian (3) undergoes an extra, finite renormalization caused by $<\!\!\varphi\!\!>_{\rm vac} \neq 0$ that may be called {\it spontaneous}. It leads from the primary bare to new bare quantities. In fact, applying the finite renormalization defined by the transformations{\footnote{Here, $F_{\mu \nu}Z^{-1/2} \rightarrow F_{\mu \nu}$ means that $F_{\mu \nu}Z^{-1/2} = F^{(\rm ren)}_{\mu \nu}$ and then the finite renormalization label (ren) is suppressed in our notation, etc.}}

\begin{equation}
 F_{\mu\,\nu} Z^{-1/2} \rightarrow F_{\mu\,\nu} \;\;,\;\; A_\mu Z^{-1/2} \rightarrow A_\mu \;\;,\;\; \psi Z^{-1/4} \rightarrow \psi \;\;,\;\; \varphi \rightarrow \varphi 
\end{equation}

\ni and

\begin{equation}
e Z^{1/2} \rightarrow e \;\;,\;\; f Z \rightarrow f \;\;,\;\; f' Z \rightarrow f' \;,
\end{equation}

\ni we transit from the primary  Lagrangian (3) to an equal new Lagrangian:

\begin{eqnarray}
\cal{L} \rightarrow \cal{L}\!\!\!& = &\!\!\!-\frac{1}{4} F_{\mu\,\nu} F^{\mu\,\nu} \left[1+ \frac{4f}{M^2}\left(2\!<\!\!\varphi\!\!>_{\rm vac} \varphi_{\rm ph} + \varphi^2_{\rm ph}\right)\right] \nonumber \\ 
& &\!\!\!-\frac{f'}{M^2} \left(\bar{\psi} \sigma_{\mu\,\nu}\psi \right) F^{\mu\,\nu} \left(<\!\!\varphi\!\!>_{\rm vac} + \varphi_{\rm ph} \right) - j_\mu  A^\mu \,.
\end{eqnarray}

\ni Here, a new bare electric charge $e$ in a new \SM electromagnetic current $j_\mu \propto e$ is spontaneously formed by the tranformation $e Z^{1/2} \rightarrow e$ with $j_\mu/e \rightarrow j_\mu/e$. Under this renormalization, the coupling constants $f$ and  $f'$ behave formally as $e^2$, while the fields $F_{\mu\,\nu}$ and $\psi$ follow the pattern of $e^{-1}$ and $e^{-1/2}$, respectively. Consequently, in the second Ref. [1] it is tentatively proposed that numerically $f \sim e^2/4$ and $f' \sim e^2/2$.

At the same time, for the sterino kinetic Lagrangian we get 

\begin{equation}
\bar{\psi}\left(\gamma \cdot p - m_{\rm sto} \right) \psi \rightarrow \bar{\psi}\left(\gamma\cdot p - m_{\rm sto} \right) \psi 
\end{equation}

\ni after our finite renormalization where in addition

\begin{equation}
m_{\rm sto}  Z^{1/2} \rightarrow  m_{\rm sto} \;\;,\;\; p Z^{1/2}  \rightarrow p \,.
\end{equation}

\ni Here, the primary sterino bare mass $m_{\rm sto} = Y <\!\!\varphi\!\!>_{\rm vac}$,  when multiplied by $Z^{1/2}$, goes over into a new sterino bare mass $m_{\rm sto} = Y <\!\!\varphi\!\!>_{\rm vac}$. Thus, 

\vspace{-0.2cm}

\begin{equation}
Y Z^{1/2} \rightarrow  Y\;,
\end{equation}

\ni as $<\!\!\varphi\!\!>_{\rm vac} \rightarrow <\!\!\varphi\!\!>_{\rm vac}$ due to the last Eq. (6).

Note that the mediating field $A_{\mu \nu}$, discussed later on in Section 4, does not change during the finite renormalization defined in Eqs. (6) and (7), $A_{\mu \nu} \rightarrow A_{\mu \nu}$ (similarly, the steron field does not change, $\varphi \rightarrow \varphi $, and also all \SM fields). The same is true for the corresponding mass, $M \rightarrow M$ (similarly, for the steron mass, $m_{\rm stn} \rightarrow m_{\rm stn}$, and also for all \SM masses).

The usual renormalization, as far as it works for effective interactions, goes normally, now from the new bare quantities to the fully renormalized quantities (after the full renormalization we have $e^2 \equiv 4\pi \alpha = 1/10.9 = 0.0917$ at low energies).

\vspace{0.3cm}

\ni {\bf 3. Quasi-magnetism of cold dark matter}

\vspace{0.3cm} 

The Lagrangian (8) implies the electromagnetic field equation

\vspace{-0.2cm}

\begin{equation}
\partial^\nu F_{\mu\,\nu} = -j_\mu - \delta j_\mu  
\end{equation}

\ni (with $F_{\mu\,\nu} = \partial_\mu A_\nu - \partial_\nu A_\mu $), where

\vspace{-0.2cm}

\begin{eqnarray}
\delta j_\mu & \equiv & \frac{4}{M^2}\,\partial^\nu\! \left[f F_{\mu\,\nu}\left(2<\!\!\varphi\!\!>_{\rm vac}\varphi_{\rm ph}\! + \varphi^2_{\rm ph}\right) + \frac{1}{2}f' \left(\bar\psi \sigma_{\mu\,\nu} \psi \right) 
\left(\!<\!\!\varphi\!\!>_{\rm vac}\! + \varphi_{\rm ph}\right)\right] \nonumber \\ 
&\sim & \frac{e^2}{M^2}\,\partial^\nu\! \left[\! F_{\mu\,\nu}\left(2<\!\!\varphi\!\!>_{\rm vac}\!\varphi_{\rm ph} + \varphi^2_{\rm ph}\right) +  \left(\bar\psi \sigma_{\mu\,\nu} \psi \right)\left(<\!\!\varphi\!\!>_{\rm vac}\! + \varphi_{\rm ph}\right)\right] 
\end{eqnarray}

\ni is a {\it quasi-magnetic} correction to the \SM electromagnetic current $j_\mu$, as $\partial^\mu \delta j_\mu \equiv 0$ identically, while $\partial_\mu  j^\mu = 0$ dynamically (so the correction $\delta j_\mu$ carries no gauge charge). In the second step in Eq. (13), our tentative proposal of $f \sim e^2/4$ and $f' \sim e^2/2$ is applied.

We can see from Eq. (8) that --- due to the photonic portal --- sterinos, though they are sterile, display the effective quasi-magnetic interaction

\vspace{-0.2cm}

\begin{equation} 
-\mu_{\rm eff}\left(\bar{\psi} \sigma_{\mu\,\nu}\psi \right) F^{\mu\,\nu}
\end{equation}

\ni proportional to the sterino quasi-magnetic moment

\begin{equation} 
\mu_{\rm eff} \equiv \frac{f'\!\!<\!\!\varphi\!\!>_{\rm vac}}{M^2} \sim \frac{e^2<\!\!\varphi\!\!>_{\rm vac}}{2M^2} \,,
\end{equation}

\ni where in the second step $f'\sim e^2/2$ tentatively. Thus, the cold dark matter consisting of sterinos is quasi-magnetic, what means that it can interact effectively with cosmic (and laboratory) magnetic fields! Thus, the phenomenon of polarization of cold dark matter in external magnetic fields may appear, leading to its quasi-magnetization.

\vspace{0.3cm}

\ni {\bf 4. A more fundamental level}

\vspace{0.3cm} 

Note finally that a third term $-(f''/M^2)(\bar{\psi} \sigma_{\mu\,\nu}\psi )(\bar{\psi} \sigma^{\mu\,\nu}\psi )$ might be introduced into the effective coupling (1), suggesting then a kind of sterile universality, where it would be natural to put $f : f' : f'' = 1 : 2 : 1$. Such a universality can be achievied by the conjecture that, at a more fundamental level, the sterile "tensor current" $ F_{\mu \nu} \varphi +\bar{\psi} \sigma_{\mu\,\nu}\psi $ appears to be coupled to an antisymmetric tensor field $A^{\mu \nu}$ (of dimension one) carrying a very large mass $M$ [1].

At this fundamental level, the electromagnetic Lagrangian  extended by the universal coupling of sterile particle fields to the $A^{\mu \nu}$ field as well as by the kinetic term of $A^{\mu \nu}$ field  takes the form

\begin{eqnarray}
{\cal{ L }_{\rm fund}} & = &  -\frac{1}{4}F_{\mu\,\nu} F^{\mu\,\nu} - j_\mu A^\mu - \frac{1}{4} \left[\left(\partial_\lambda A_{\mu \nu}\right)\left(\partial^\lambda A^{\mu \nu}\right) + M^2 A_{\mu \nu}A^{\mu \nu}\right] \nonumber \\ & &- \sqrt{f}\left(F_{\mu \nu}\varphi + \bar\psi \sigma_{\mu\,\nu} \psi \right) A^{\mu \nu} \;, 
\end{eqnarray}

\ni if $f : f' : f'' = 1 : 2 : 1$ ({\it cf.} the second Ref. [1]). It implies the following two field equations for $ F_{\mu \nu}$ and $ A_{\mu \nu}$:

\begin{equation}
\partial^\nu \left( F_{\mu\,\nu} + 2 \sqrt{f}\, \varphi A_{\mu \nu} \right) = - j_\mu \;\;,\;\; \left( \Box - M^2\right) A_{\mu \nu} = - 2\sqrt{f}\left(F_{\mu \nu}\varphi + \bar\psi \sigma_{\mu \nu} \psi \right)  
\end{equation}

\ni (with $F_{\mu\,\nu} = \partial_\mu A_\nu - \partial_\nu A_\mu $ and $\partial_\nu\partial^\nu = - \Box$). The first Eq. (17) can be put in the form

\begin{equation}
\partial^\nu F_{\mu\,\nu}  = - j_\mu - \delta j_\mu \;\;{\rm or}\;\; \partial^\nu\!\left( F_{\mu \nu} + \delta F_{\mu \nu}\right) = - j_\mu  \;,
\end{equation}

\ni where

\vspace{-0.2cm}

\begin{equation}
\delta j_\mu  \equiv 2 \sqrt{f}\, \partial^\nu \left( \varphi A_{\mu \nu} \right) \equiv  \partial^\nu \delta F_{\mu \nu} \;\;,\;\; \delta F_{\mu \nu} \equiv 2\sqrt{f}\, \varphi\, A_{\mu \nu} \;\;,\;\; \partial^\nu \delta j_\mu \equiv 0 \;. 
\end{equation}

If the momentum transfers through the field $A_{\mu \nu}$ are approximately negligible {\it versus} $M^2$, then from the second Eq. (17)

\vspace{-0.2cm}

\begin{equation}
A_{\mu \nu} \simeq  \frac{2\sqrt{f}}{M^2}\left( F_{\mu\,\nu}\varphi + \bar\psi \sigma_{\mu \nu} \psi \right)  
\end{equation}

\ni and so, from the first Eq. (19) and the identity $\varphi \equiv <\!\!\varphi\!\!>_{\rm vac} +\varphi_{\rm ph}$ 

\begin{eqnarray}
\delta j_\mu & \!\!\!\simeq\!\!\!\! & \;\;\;\frac{4f}{M^2}\,\partial^\nu\!\left[\varphi (F_{\mu\,\nu} \varphi  + \bar\psi \sigma_{\mu\,\nu} \psi)\right] \equiv \frac{4f}{M^2} \partial^\nu\!F_{\mu\,\nu} <\!\!\varphi\!\!>^2_{\rm vac} \nonumber\\
 &  & + \frac{4f}{M^2} \partial^\nu\!\left[F_{\mu\,\nu}\left(2\!<\!\!\varphi\!\!>_{\rm vac} \varphi_{\rm ph} + \varphi^2_{\rm ph}\right) + (\bar\psi \sigma_{\mu\,\nu} \psi)\left(<\!\!\varphi\!\!>_{\rm vac} + \varphi_{\rm ph}\right)\right]. 
\end{eqnarray}

\vspace{0.1cm}

\ni Hence, using the first Eq. (18) and the definition (5) of $Z^{-1}$, we obtain (multiplying by $Z^{1/2}$):

\begin{equation}
\partial^\nu\! F_{\mu\,\nu} Z^{-1/2}\! \simeq -j_\mu Z^{1/2}\! - \frac{4f Z}{M^2} \partial^\nu\!\left[F_{\mu\,\nu}\!\left(2\!<\!\!\varphi\!\!>_{\rm vac} \varphi_{\rm ph} \!+\! \varphi^2_{\rm ph}\right) \!+\! (\bar\psi \sigma_{\mu\,\nu} \psi)(\!<\!\!\varphi\!\!>_{\rm vac} \!+\! \varphi_{\rm ph}\!)\right]\! Z^{-1/2}. 
\end{equation}

\vspace{0.1cm}

\ni After the finite renormalization (6) and (7), this gives $\partial^\nu\! F_{\mu\,\nu} \simeq - j_\mu - \delta j_\mu $ (see Eq. (12)) with a new $\delta j_\mu$ as calculated in Eq. (13) from the effective Lagrangian (8) (valid after this finite renormalization). Here, $f = f'/2$.

\vspace{0.3cm}

\ni {\bf 5. Final remarks}

\vspace{0.3cm} 

In the case of effective interaction (1) (when our effective Lagrangian (8) works) it is shown in the second Ref. [1] that sterinos --- as candidates for cold dark matter --- are consistent with its observed relic abundance $\Omega_{\rm DM} h^2 \sim 0.1$ [3], if they freeze out thermally in the early Universe with the mass

\begin{equation} 
m_{\rm sto} \sim 0.6 \;{\rm TeV} \,.
\end{equation}

\ni This estimation is valid under the tentative assumptions that the sterino and steron masses as well as the mass scale $M$ are similar,

\begin{equation} 
m_{\rm sto} \sim m_{\rm stn} \sim M \,,
\end{equation}

\ni and that the magnitude of coupling constant $f' = 2f$ is near to $e^2/2$,

\begin{equation} 
 f' = 2f \sim \frac{1}{2} e^2 \,.
\end{equation}

\ni In this case, sterons are unstable at the Universe time-scale.

Due to the quasi-magnetism of sterinos, as it is described by the interaction (14), they could be produced in the high energy collisions of \SM particles {\it via} virtual photons. The simplest production process of this kind would be

\begin{equation}
e^+ \, e^- \rightarrow \gamma^* \rightarrow  ({\rm antisterino})({\rm sterino}) 
\end{equation}

\ni with the centre-of-mass threshold energy $\sim$ 1.2 TeV if our tentative estimate of the sterino mass $m_{\rm sto}$ was applied.

Making use of the coupling

\begin{equation}  
e\bar{\psi}_e \gamma_\mu \psi_e A^\mu - \mu_{\rm eff}\,\bar{\psi} \sigma_{\mu \nu} \psi  F^{\mu \nu} 
\end{equation}

\ni (see Eq. (14)), where $e = |e|$ and $\mu_{\rm eff} \equiv f' <\!\!\varphi\!\!>_{\rm vac}/M^2$, while $\psi_e$ and $\psi$ are the electron and sterino fields, we can calculate in a conventional way [4] the following differential and total cross-sections for the sterino-antisterino production process (26) in the electron-positron 
centre-of-mass frame:

\begin{equation}
\!\!\!\frac{d \sigma}{d \Omega_{\rm sto}} (e^+ \, e^- \!\rightarrow {\rm asto\;\rm sto}) = \frac{1}{2v_e} \frac{1}{(2\pi)^2} \!\left(\!\frac{e f' \!<\!\!\varphi\!\!>_{\rm vac}}{M^2}\right)^{\!\!2}\!\!\frac{|{\vec{p}_e|}}{2E_e } \!\!\left[1 \!+\! \frac{m^2_{\rm sto}}{E^2_e} \!- \!\left(1 \!-\! \frac{m^2_{\rm sto}}{E^2_e}\right)\!\cos^2 \!\theta_{\rm sto}\!\right]
\end{equation}

\ni and

\begin{equation}
\sigma(e^+ \, e^- \!\rightarrow {\rm asto\;\rm sto}) 2v_e =   \frac{1}{3\pi} \left(\frac{e f' \!<\!\!\varphi\!\!>_{\rm vac}}{M^2} \right)^{\!\!2}\left(1 + 2\frac{m^2_{\rm sto}}{E^2_e}\right) \,.
\end{equation}

\vspace{0.2cm}

\ni Here, ${\vec{p}}_{\rm sto}\cdot{\vec{p}}_e = |{\vec{p}}_{\rm sto}||{\vec{p}}_e| \cos \theta_{\rm sto}$,  $d\Omega_{\rm sto} = 2\pi \sin \theta_{\rm sto} d\theta_{\rm sto}$ and  $v_e = |{\vec{p_e}}|/E_e $. In Eqs. (28) and (29), the electron mass $m_e$ is negligible due to the condition of $m^2_e/E^2_e  \ll 1$. In the case of our tentative assumptions of $m_{\rm sto} \sim m_{\rm stn} \sim M$ and $f' \sim e^2/2$, we get $m_{\rm sto} \sim 0.6 \;{\rm TeV}$ (after the full renormalization we have $e^2 = 1/10.9 = 0.0917$ at low energies).

Our concluding sentences may read as follows. The presented model of cold dark matter is based on two sterile fields $\psi$ and $\varphi$ (sterinos and sterons) and one mediating field $A_{\mu \nu}$. All are coupled to the electromagnetic field $F_{\mu \nu}$ forming then the photonic portal from the \SM world to sterile world, the latter displaying in such a case the quasi-magnetism. This unifies the \SM electrodynamics with the dynamics of sterile world (see Eqs. (17) and also (18)). Such a construction works after the \SM electroweak symmetry is broken and photons  emerge. Thus, the new unification might be dual or complementary to the previous electroweak unification ({\it contraria sunt complementa} --- Niels Bohr).

The enthusiasts of supersymmetry are right to emphasize the fact that the experimental estimation $\Omega_{\rm DM} h^2 \sim 0.1$ --- implying for a particle-antiparticle pair of cold dark matter (possibly of Majorana character) the total annihilation cross-section $\sim 1$ pb --- is consistent with neutralinos $\chi$ as candidates for the thermal dark matter if  $m_\chi \sim 0.1$ TeV. However, our sterinos can be also consistent with the experimental value of $\Omega_{\rm DM} h^2 \sim 0.1$ if $m_{\rm sto} \sim 0.6$ TeV (as is shown under the tentative assumptions of  $m_{\rm sto} \sim m_{\rm stn} \sim M$ and $f \sim e^2/4$). This qualifies also sterinos as candidates for the thermal dark matter.  
 

\vspace{0.6cm}

{\centerline{\bf References}}

\vspace{0.4cm}

{\everypar={\hangindent=0.65truecm}
\parindent=0pt\frenchspacing

{\everypar={\hangindent=0.65truecm}
\parindent=0pt\frenchspacing

~[1]~W.~Kr\'{o}likowski, arXiv: 0712.0505 [{\tt hep--ph}]; arXiv: 0803.2977 [{\tt hep--ph}].

\vspace{0.2cm}

~[2]~J. March-Russell, S.M. West, D. Cumberbath and D.~Hooper, arXiv: 0801.3440v2 [{\tt hep-ph}].

\vspace{0.2cm}

~[3]~For recent reviews {\it cf.} G. Bartone, D.~Hooper and J.~Silk, {\it Phys. Rept.} {\bf 405}, 279 (2005); M.~Taoso, G.~Bartone and A.~Masiero, arXiv: 0711.4996 [{\tt astro-ph}]; {\it cf.} also E.W.~Kolb and S.~Turner, {\it Early Universe} (Addison-Wesley, Reading, Mass., 1994); and K.~Griest and D.~Seckel, {\it Phys. Rev.} {\bf D 43}, 3191 (1991). 

\vspace{0.2cm}

~[4]~J.D.~Bjorken and S.D.~Drell, {\it Relativistic Quantum Mechanics} (McGraw-Hill, New York, 1964). 

\vspace{0.2cm}

\vfill\eject

\end{document}